\documentclass[12pt]{article}

\usepackage{authblk}
\usepackage{amsmath}
\usepackage{mathtools}
\usepackage{amssymb} 
\usepackage{amsthm}
\usepackage{bbm}
\usepackage{graphicx,epsf}
\usepackage{epstopdf}
\usepackage{enumerate}
\usepackage{caption}
\usepackage{natbib}
\usepackage{bm}
\usepackage{xcolor}
\usepackage{tabularx}

\newtheorem{proposition}{Proposition}

\newtheorem{corollary}{Corollary}
\theoremstyle{definition}

\newcommand{\sgn}{\mathrm{sgn}}
\newcommand{\Exp}{\mathrm{E}}
\newcommand{\med}{\mbox{median}}

\newcommand{\bx}{\bm x}
\newcommand{\by}{\bm y}
\newcommand{\lb}{\theta_{\mbox{\tiny lo}}} 
\newcommand{\ub}{\theta_{\mbox{\tiny hi}}} 
\newcommand{\betah}{\hat{\beta}}

\author{Jakob Raymaekers$^{1}$ \and Florian Dufey$^{2}$}
\date{%
    $^1$ Department of Quantitative Economics\\
Maastricht University\\
Maastricht, The Netherlands\\
    $^2$ Roche Diagnostics GmbH\\
		Nonnenwald 2, 82377 Penzberg\\
		Germany.
}

\begin{document}
\title{\Large Equivariant Passing--Bablok regression in quasilinear time}
\maketitle

\begin{abstract}
\noindent Passing--Bablok regression is a standard tool for method and assay comparison studies thanks to its place in industry guidelines such as CLSI. Unfortunately, its computational cost is high as a naive approach requires $\mathcal{O}(n^2)$ time. This makes it impossible to compute the Passing--Bablok regression estimator on large datasets. Additionally, even on smaller datasets it can be difficult to perform bootstrap-based inference. We introduce the first quasilinear time algorithm for the equivariant Passing--Bablok estimator. In contrast to the naive algorithm, our algorithm runs in  $\mathcal{O}(n\log(n))$ expected time using $\mathcal{O}(n)$ space, allowing for its application to much larger data sets. Additionally, we introduce a fast estimator for the variance of the Passing--Bablok slope and discuss statistical inference based on bootstrap and this variance estimate. Finally, we propose a diagnostic plot to identify influential points in Passing--Bablok regression. The superior performance of the proposed methods is illustrated on real data examples of clinical method comparison studies. 
\end{abstract}

\clearpage

\section{Introduction}

\citet{passing1983new} introduced their ``shifted median'' algorithm for robust linear regression with measurement error\footnote{With ``measurement error'' we mean the situation where not only the  abscissa values but also the ordinate values have been measured with error. In this situation, standard methods, like linear regression, yield biased estimates for the slope and intercept.}, a situation typically encountered in clinical biochemistry when comparing different test versions, instruments or systems. 
Since its introduction, the method has been widely adopted and recognized in pertinent guidelines  \citep{budd2013measurement,capiau2019official} as a de-facto standard for method and assay comparison studies.
Interesting recent applications comprise the comparison of SARS-CoV-2 antibody tests \citep{perkmann2021anti}, smartphone compatible HIV testing \citep{kanakasabapathy2017rapid} or the calibration of tumor mutational burden in lung cancer  with genomic assays \citep{chang2019bioinformatic} to name just a few.  
Potential alternative methods, like major axis (Deming) regression or reduced mayor axis (geometric mean) regression suffer from non-robustness against outlying measurements. 

Passing and Bablok justified the algorithm by giving mainly heuristic arguments, like its similarity to Theil--Sen regression and its invariance under an interchange of ordinate and abscissa. 
By design, the original method was restricted to expected slopes near 1, the typical situation when comparing measurement methods believed to be equivalent. 
In a latter paper \citet{bablok1988general} explored several possibilities to transcend this limitation, to the regime of ``method transformation'', as they called it. 
These kind of problems arise typically in the process of calibration, e.g.\ calibration of a new test which yields fluorescence intensity readings on an established test which yields concentrations. It is clear that intensity and concentration may be on numerically vastly different scales, so that the slope can be far from one. 

Recently \citep{dufey2020derivation}, it was shown how one of these methods, named ``equivariant Passing--Bablok regression (ePBr)'', can be derived by the requirement that Kendall's correlation is zero after a scaling and rotation of the data involving the slope estimate. Also improved expressions for the variances of the slope and intercept could be derived. 
In this respect, ePBr is very similar to reduced major axis regression, which can be obtained  replacing Kendall correlation in the procedure described above with Pearson correlation.
Yet another characterization of the ePBR slope is that it is equal to the median of the absolute slopes of the lines through all pairs of observations. This last definition of ePBR is most commonly used for its computation. More precisely, the ePBR is typically computed by taking the median of all $n (n-1)/2$  absolute slopes formed by the lines through each pair of data points (where $n$ denotes the number of observations). Clearly, this naive method of computation requires $\mathcal{O}(n^2)$ time which quickly becomes infeasible for large data sets. 

We introduce the first quasi-linear time algorithm for the equivariant Passing--Bablok estimator. 
More specifically, our algorithm runs in $\mathcal{O}(n\log(n))$ expected time using $\mathcal{O}(n)$ space and is guaranteed to yield the ePB regression estimate. 
Our algorithm is based on the ideas of slope selection algorithms for the Theil--Sen \citep{Theil1950,sen1968estimates} and repeated median estimators \citep{siegel1982robust}. 
In particular, we construct a randomized algorithm \citep{motwani1995randomized} much in the style of  \cite{mount1991computationally, matouvsek1991randomized,dillencourt1992randomized, matouvsek1998efficient} and prove its computational complexity. In addition to a fast algorithm for ePBr, we introduce a measure of influence which quantifies how strongly each data point influences the estimated slope and can thus be used to identify influential points in diagnostic plots. The proposed algorithm as well as the diagnostic plot are illustrated in the analysis of real data sets on the comparison of clinical tests.\par
The remainder of the paper is organized as follows. Section \ref{sec:methodology} reviews the ePB estimator and describes the proposed algorithm together with an analysis of its computational complexity. Section \ref{sec:simulation} presents a simulation study comparing the computation times of the proposed algorithm with the commonly used naive approach. An application of the proposal to real data is shown in Section \ref{sec:examples}. Finally, Section \ref{sec:conclusion} concludes.

\section{Methodology}\label{sec:methodology}

\subsection{Equivariant Passing--Bablok regression}
We start by laying out the assumptions of the underlying regression model. Suppose we observe 2 sequences of random variables $X_i$ and $Y_i$ of the form $X_i = \tilde{X}_i + \xi_i$ and $Y_i = \tilde{Y}_i + \eta_i$, where $\xi_i$ and $\eta_i$ denote the error terms. Furthermore, assume that the error terms satisfy:
\begin{itemize}
    \item For all $i\neq j$, the error terms are independent.
    \item For all $i$, $\xi_i$ and $\eta_i$ are independent.
    \item $\Exp[\xi_i] = \Exp[\eta_i] = 0$ for all $i$ and $\beta^2 \coloneqq \sigma_{\xi_i}^2/\sigma_{\eta_i}^2$ is constant.
    \item For fixed $i$, the distributions of $\xi_i / \sigma_{\xi_i}$ and $\eta_i/\sigma_{\eta_i}$ are equal.
\end{itemize}

Under these conditions, we assume a linear relation between the expected values of two sequences of random variables $X_i$ and $Y_i$ of the form 
\begin{equation}\label{eq:pb}
    \tilde{Y}_i = \alpha + \beta \tilde{X}_i
\end{equation}
Like in reduced major axis regression, it is assumed that the parameter $\beta$ describes both the expected slope and the ratio of the error's standard deviations. 
Note that these conditions are not very stringent. In particular, neither normality nor homoscedasticity of the error terms is assumed.  Additionally, in contrast to ordinary least squares  or Theil--Sen regression, it is not assumed that the $X_i$ are free of error. 

We are now interested in estimating the intercept and slope parameters $\alpha$ and $\beta$ of Eq.\ \ref{eq:pb}. We assume having observed a sample $(\bx,\by)$ of $n$ data points $(x_i, y_i)$ with $i = 1,\ldots,n$. 
The equivariant Passing--Bablok (ePB) regression, introduced by \cite{bablok1988general}, is equivalent to finding the median of the absolute value of the pairwise slopes \citep{dufey2020derivation}:

\begin{equation*}
\hat{\beta} = \med_{i < j}{\left|s_{ij}\right|},
\end{equation*} 
where $s_{ij}=\frac{y_j - y_i}{x_j - x_i}$. Quite evidently, this estimate can be calculated by 
enumerating all pairwise absolute slopes and calculating the median. However, this naive approach has a computational complexity of $\mathcal{O}(n^2)$ which quickly becomes prohibitive on large data sets.

\subsection{Passing--Bablok regression in quasilinear time}
We now introduce an algorithm for computing the ePB regression estimator in quasilinear time. 
We will first describe the algorithm assuming that all data points are distinct. Afterwards, the treatment of ties in the data as well as degeneracies in the algorithm are discussed.\\

\subsubsection{Description of the base algorithm}
Assume we are given $n$ data points $(x_i, y_i)$ with $i = 1,\ldots,n$. 
Our goal is to find the median of the absolute values of the slopes formed by connecting two data points. For simplicity, we assume here that there are no duplicate values in the dataset, i.e. \ $x_i \neq x_j$ and $y_i \neq y_j$ for all $i \neq j$ with $i,j\in \{1, \ldots, n\}$. We now look for the $k$th absolute slope, where
\begin{equation}\label{eq:uppermedian}
k =
\begin{cases}
\frac{K}{2} + 1 \mbox{ if } K \mbox{ is even}\\
\frac{K+1}{2} \mbox{ if } K \mbox{ is odd}\\
\end{cases}
\end{equation} with $K = \frac{n(n-1)}{2}$. This is also known as the ``upper median''. 
Note that the exact same algorithm can be used for any other order statistic of the absolute slopes by appropriately adjusting $k$. 
The case of duplicate values will be addressed in the next section.\\

Our randomized algorithm builds on the following dual representation of the data points suggested by \cite{cole1989optimal}. 
To each data point $(x_i, y_i)$ we associate a line in dual space given by
$$l_i:v =  y_i-x_i u.$$

This association may be interpreted in the original space: Consider a fixed $(u,v)$, then we have that the line through $(x_i,y_i)$ with slope $u$ in the original space has intercept $v$.\par Now it is critical to recognize that the slope $s_{ij}$ is equal to the u-coordinate of the intersection of $l_i$ and $l_j$ in dual space. Therefore, finding the median absolute slope is equal to finding the median of the absolute intersection abscissas (IAs) in dual space. This idea is illustrated in Figure \ref{fig:Fig1} below.
This figure shows the data points $\{(0.5,2),(1,3),(2, 1)\}$ in primal space (left panel) and as lines in dual space (right panel). As an example, we have $s_{12} =\frac{3 - 1}{1 - 0.5}= 2$, and this corresponds with the $u$-coordinate of the intersection of $l_1: v = 2 - 0.5u$ and $l_2: v = 3 - u$ in dual space. 
\begin{figure}[!ht]
    \centering
    \includegraphics[width = \linewidth]{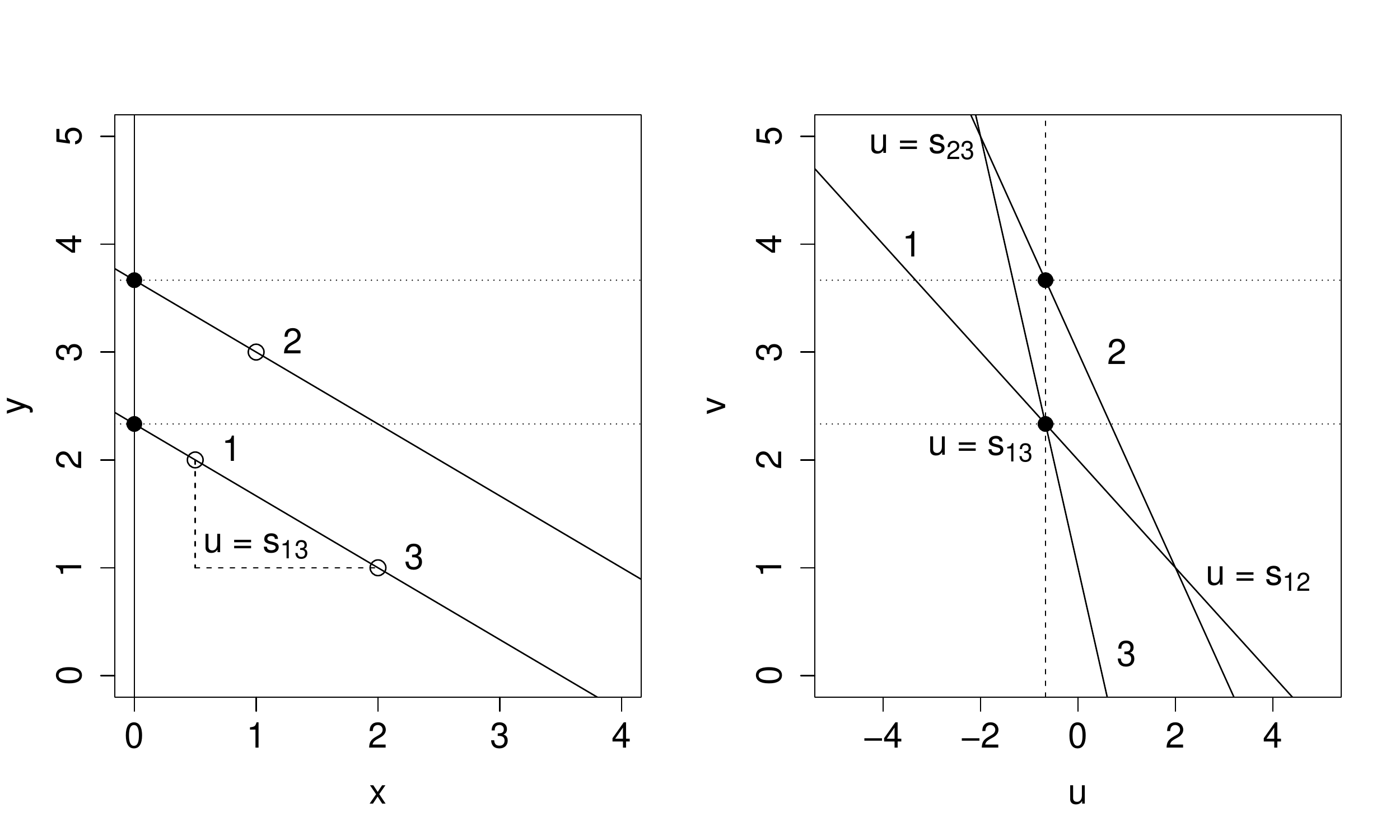}
     \captionsetup{width=\linewidth}
    \caption{\label{fig:Fig1}Representation of the data points $\{(0.5,2),(1,3),(2, 1)\}$ in primal space (left) and as lines in dual space (right). Shown are also lines with slope $u$ (here, $u=s_{13}$ (dashed) as an example) passing through the points 1 -- 3 in primal space. The intercepts $v$ with the y-axis are plotted as a function of $u$ in dual space (cf.\ the dotted horizontal lines). As the line with slope $u=s_{13}$ passes through both points 1 and 3, their intercept coincides, so that the common line is matched to a point in dual space. }
\end{figure}

This representation of the data points lies at the basis of the following result which is used and proved in different forms in \cite{mount1991computationally, matouvsek1991randomized,dillencourt1992randomized}. 

\begin{proposition}\label{prop:0}
For any interval $(a,b]$, denote $I = \{s_{ij}|s_{ij} \in (a,b]\}$. 
We then have
\begin{enumerate}
\item The number of slopes in $(a,b]$ can be counted in $\mathcal{O}(n\log(n))$ time.
\item The slopes in $(a,b]$ can be enumerated in $\mathcal{O}(n\log(n) + |I|)$ time.
\item We can sample $q$ elements out of $I$ with replacement in $\mathcal{O}(n\log(n) + q)$ time.
\end{enumerate}
\end{proposition}
All three of the tasks in the proposition can be solved using the idea of sampling and counting inversions in a vector. An inversion in a vector $z_1, \ldots, z_n$ is a pair of indices $i$, $j$ for which $i < j$ and $z_i > z_j$.
Counting and sampling inversions can be achieved by using modifications of the merge sort algorithm, which has a quasilinear time complexity. For the purpose of illustration, we give the following example of the relation between counting slopes and inversions. We already discussed that counting the number of slopes in an interval is equivalent to counting the number of IAs in the same interval in dual space. Now, this is equivalent to counting the number of inversions in the permutation obtained by going from the order of the intersections of the lines with the vertical line $u = a$ to the order of the intersections of the lines with the vertical line $u = b$. Counting the number of inversions in a vector can be done by an adapted merge sort algorithm, proving the first statement of the proposition above.\\

The main idea is now to maintain a collection of two intervals which is guaranteed to contain the median absolute slope, and make this set progressively smaller until it is small enough to inspect each of its elements and select the desired absolute slope. 
In order to do this, we maintain two non-negative real numbers $\lb$ and $\ub$ with $\lb < \ub$ so that the target slope is always contained in the set $[-\ub, -\lb) \cup (\lb, \ub]$. Denote with $\Theta$ the set of all absolute slopes contained in this set and $|\Theta|$ the number of elements in $\Theta$. We also denote with $C_l$, $C_m$ and $C_h$ the number of slopes in $[-\ub, -\lb)$, $[-\lb, \lb] $ and $(\lb, \ub]$ respectively. 
Note that $|\Theta| = C_l + C_h$. Given a certain $\Theta$, we are looking for the $k^*$-th smallest absolute value in $\Theta$ where $\Theta$ where $k^* = k - C_m$. 
The initial values of these parameters are given by $\lb = 0$, $\ub = \infty$, $C_l = |\{s_{ij}|s_{ij} \in (-\infty,0)\}|$, $C_l = |\{s_{ij}|s_{ij} \in (0,\infty)\}|$, and $C_m = |\{s_{ij}|s_{ij} =0\}|$.\\

Now we use the following extension of Proposition \ref{prop:0} (for a proof we refer to the supplementary material):
\begin{proposition}\label{prop:1}
Under the notation above, we have:
\begin{enumerate}
\item The number of absolute slopes in any interval $(a,b]$ can be counted in $\mathcal{O}(n\log(n))$ time.
\item The absolute slopes in $\Theta$ can be enumerated in $\mathcal{O}(n\log(n) + |\Theta|)$ time.
\item Given fixed $\lb$ and $\ub$, we can sample $q$ elements out of $\Theta$ with replacement in $\mathcal{O}(n\log(n) + q)$ time.
\end{enumerate}
\end{proposition}

From the second statement in this proposition we obtain the following corollary, stating that if the number of slopes in $\Theta$ is small enough, we can solve the problem in $\mathcal{O}(n\log(n))$ time by enumerating all of them and selecting the target slope. 
\begin{corollary}\label{col:1}
Given $|\Theta| = \mathcal{O}(n)$, we can find any order statistic of the elements in $\Theta$ in $\mathcal{O}(n\log(n))$ time.
\end{corollary}

In each iteration of the algorithm, we thus start by checking whether $|\Theta| < c$ where $c = \mathcal{O}(n)$. 
If this condition is satisfied, we invoke Corollary \ref{col:1} and terminate the algorithm. 
In case $|\Theta|$ is too large, we cannot resort to the corollary above and therefore, we build a strategy for contracting the intervals by making $|\ub - \lb|$ progressively smaller. 
We start by randomly sampling $n$ elements from $\Theta$, which can be done in $\mathcal{O}(n\log(n))$ time using Proposition \ref{prop:1}. Denote the vector of absolute values of these elements as $S$.
For a $t > 0$ (see later for the specific value), we then compute 
 \begin{align*}
k_l^* = \max\left(1, \left\lfloor\frac{k^*}{|\Theta|}n - t\frac{\sqrt{n}}{2}\right\rfloor\right)\\
k_h^* = \min\left(r,\left\lceil \frac{k^*}{|\Theta|}n + t\frac{\sqrt{n}}{2}\right\rceil\right)\\
\end{align*}
and the associated candidate lower and upper bounds as the corresponding order statistics of the absolute slopes:
 \begin{align*}
\lb' &= S_{(k_l^*)}\\
\ub' &= S_{(k_h^*)}.
\end{align*}
Now we count the number of absolute slopes in $(\lb, \lb']$, call this $c_1$, as well as the number of absolute slopes in $(\lb', \ub']$, call this $c_2$.\\

Using these counts, we can update the numbers $\lb$ and $\ub$ as follows. If $c_1\geq k^*$, we know that the $k^*$-th absolute slope in $\Theta$ must lie in the interval $(\lb, \lb']$. In that case, we can update $\ub \leftarrow \lb'$. If, on the other hand, we have $c_1< k$ and $c_1 + c_2 \geq k^*$, we know that the $k^*$-th absolute slope of $\Theta$ must be in $(\lb', \ub']$, hence we can update $\lb \leftarrow \lb'$ and $\ub \leftarrow \ub'$. 
If none of these conditions hold, we know that the target slope must be larger than $\lb'$ in absolute value, and we thus update $\lb \leftarrow \ub'$. We have now made $|\ub - \lb|$  smaller, and we can update the counts $C_l$, $C_m$ and $C_h$ as well as $k^*$ before starting the next iteration. The iterative procedure stops when $|\Theta| = \mathcal{O}(n)$, after which we invoke Corollary \ref{col:1}.\\ 

We can now state the expected run time of the algorithm. For a proof we refer to the supplementary material.

\begin{proposition}\label{prop:2}
The proposed algorithm with properly chosen values for $t$ and $c$ runs in an expected quasilinear time of $\mathcal{O}(n\log(n))$ and uses $\mathcal{O}(n)$ space.
\end{proposition}

As detailed in the supplementary material, there are multiple ways of choosing $t$ and $c$ in order to meet the conditions of Proposition \ref{prop:2}. In our implementation of the algorithm, we use $t = 3$ and $c = 20$.

\subsubsection{Treatment of duplicate values}
In the above explanation we have assumed that all $x$ and $y$ values are distinct. 
This guarantees that we do not have any slopes equal to zero, nor any undefined slopes (equal to $\infty$). 
In practice however, duplicate values are quite common and thus they need to be dealt with appropriately. 
There are three types of duplicate values: $x$-only, $y$-only and $xy$ duplicates. 
The first type are observations with the same $x$ value, but different $y$-values.
The second type are observations with the same $y$ value, but different $x$-values.
The third type are observations with the same values for $x$ and $y$.
The correct treatment of these ties can be inferred from the more fundamental expression for the PB estimator \citep{dufey2020derivation}  which results as a solution of 
\begin{equation}\label{eq:PB}
    \sum_{i<j} \sgn(\Delta y_{ij}+\Hat{\beta}\Delta x_{ij})\sgn(\Delta y_{ij}-\Hat{\beta}\Delta x_{ij})=0, 
\end{equation}
where $\Delta x_{ij} \coloneqq x_j - x_i$ and $\Delta y_{ij} \coloneqq y_j - y_i$.
From this formula, we see that an $x$-only duplicate generates a 1 in the sum on the left hand side. 
This will thus induce an upward bias on $\betah$ since the left hand side becomes smaller as $\betah$ increases and larger as $\betah$ gets closer to 0. 
Similarly, a $y$-only duplicate generates a $-1$ on the left hand side, and thus induces a downward bias on the estimated $\beta$. In our algorithm, we deal with this by mapping the observations with $x$-only ties to lines meeting at $+\infty$ in dual space (irrespective of the sign of the numerator of this slope), whereas the $y$-only ties are mapped to lines meeting at $0$. 
This strategy achieves the desired treatment of these duplicate values without any further adaptations to the described algorithm.\par
The $xy$-ties are different, and should be treated more carefully. First note that these ties yield zeroes on the left hand side of Equation \ref{eq:PB}, which suggests that they should be ignored in the calculation of the slope estimate. In the naive algorithm, this is typically done by removing these ``undefined'' slopes connected to such duplicate observations before starting the calculation of the median absolute slope. Unfortunately, this strategy cannot be applied for the proposed fast algorithm as we do not actually list all slopes. Instead, we map these ties to lines which meet at $\infty$ in dual space, and then adjust the target order statistic $k$ such that the overall procedure is equivalent to removing these slopes before calculating the upper median. More precisely, by mapping the $xy$-ties to lines meeting at $\infty$ in dual space, we induce an upward bias on the calculated absolute slope. As that is not the desired result, we have to adjust the target order statistic as follows. We start by counting the number of $xy$-ties. More precisely, let $T_i$ be the number of times observation $(x_i,y_i)$ occurs in the data. We can then adjust the target order statistic for the median absolute slope by setting $K=n(n-1)/2-\sum_{i=1}^n{\frac{T_i - 1}{2}}$, and then using the formula of Equation \ref{eq:uppermedian} to obtain $k$. In this way, we obtain the same result that you would obtain by removing the undefined slopes and computing the median absolute slope on the remaning values. Note that the number of $xy$-ties can be computed in $\mathcal{O}(n\log(n))$ time such that this procedure does not affect the overall running time of the algorithm.

\subsubsection{Treatment of degeneracies}
Much like other geometric algorithms, the proposed algorithm requires a careful treatment of some degenerate cases which may occur. 
To be precise, three types of degeneracies may occur:

\begin{enumerate}
\item We can have ties when sorting the intersections of the lines with vertical lines in dual space. 
\item Multiple occurrence of the same slope in $\Theta$ can give a high probability of not improving on $\ub$, i.e.\ of obtaining $\ub' = \ub$.
\item We can have (many) slopes equal to 0.
\end{enumerate}

The first two types of degeneracies are addressed in similar spirit to the strategy of \cite{dillencourt1992randomized}. 
More specifically, ties when sorting sampled slopes are solved by decreasing slope in dual space, i.e.\ by decreasing $-x_i$. 
Additionally, we use a stable sort to make sure that duplicate values do not cause random inversions. 
For the second type of degeneracy, we slightly alter the counting procedure after obtaining new candidate bounds $\lb'$ and $\ub'$. 
Instead of using the counts $c_1$ and $c_2$ as before, we now put $c_2$ equal to the number of absolute slopes in $(\lb', \ub')$ and generate an additional count $c_3$ equal to the number of absolute slopes equal to $\ub'$. 
If now $c_1 + c_2 \geq k^*$, we know there is no issue at the boundary and we can concentrate on the open interval $(\lb', \ub')$. 
If $c_1 + c_2 < k^*$ but $c_1 + c_2 + c_3 \geq k^*$, then the algorithm ends and the median absolute slope is equal to $\ub'$. 
This split of the counts $c_2$ and $c_3$ can be generated without adding to the computational complexity by slightly modifying the tie-breaking rules in the sorting of the intersections in dual space.  \\
For the third degeneracy, we start the algorithm by counting the number of slopes  equal to 0 (note that a slope of $-\infty$ never occurs since all y-ties are mapped to $\infty$). If this last number is larger than $k$, we can stop here and return 0. 
This only occurs when there are a high number of tied values in $y$ which cause the zero slopes. These initial computations also allow us to properly initialize the counts $C_l, C_m$, and $C_h$, as well as $k^*$.

\subsection{A diagnostic tool}
One of the reasons for the popularity of Passing--Bablok regression is that it is more robust to outlying observations than many of its alternatives. In case outliers are present in the sample, it is often interesting to identify them for further inspection. In the classical setting, the most detrimental outliers are those which are also leverage points. These are observations which have a large influence on the estimate and pull it away from its optimal value \citep{rousseeuw2005robust}. Note that it can be hard to identify these points if one does not use a robust estimator due to masking and swamping effects. In our setting however, we do use a robust estimator and so we can reliably consider the estimated line to be a reasonable estimate (assuming the percentage of outliers is not too large, i.e.\ smaller than $1 - 1/\sqrt{2} \approx 29 \%$, which is the breakdown value of our estimator). \\

In the setting of ePBr, we propose to assign an influence score to each point $i$ defined by an estimate of the quantile shift induced by leaving out point $i$ from the estimation of the slope. This measure is similar in spirit to the DFBETA measure \citep{belsley2005regression} in ordinary least squares regression.
More precisely, we propose to use:

\begin{equation*}
    \tau_i \coloneqq F_{n,S}(\Hat{\beta}) -  F_{n,S}(\Hat{\beta}_{-i}) = \frac{1}{2}-  F_{n,S}(\Hat{\beta}_{-i})
\end{equation*}
where $F_{n,S}$ is the empirical cumulative distribution function of the absolute slopes and $\Hat{\beta}_{-i}$ is the median absolute slope estimate when observation $i$ is left out of the sample.
Clearly, if leaving out point $i$ from the sample does not influence the estimate $\Hat{\beta}$ at all, we have $\tau_i = 0$. On the other hand, if $\tau_i\gg0$, it means that leaving out point $i$ from the sample yields an estimated slope much smaller than $\Hat{\beta}$, indicating that point $i$ pulls the estimate upwards. We can interpret $\tau_i \ll 0$ in similar fashion.

Now, one can derive that
\begin{equation*}
    \tau_i = \frac{\#\{ j \neq i \; | \;|s_{ij}|> \Hat{\beta}\} -  \#\{ j \neq i \; | \;|s_{ij}|< \Hat{\beta}\}}{n(n-1)/2}.
\end{equation*}
In other words, $\tau_i$ is proportional to the difference of the number of slopes connected with point $i$ larger than the estimated slope $\Hat{\beta}$, and the number of slopes smaller than $\Hat{\beta}$. In order to estimate this quantity, we note that $ \tau_i =\frac{\sum_{j\neq i}{\tau_{ij}}}{n(n-1)/2}$ where $\tau_{ij} = \sgn(|s_{ij}| - \Hat{\beta})$. Furthermore, we have that 
\begin{align*}
\tau_{ij}=&\;\sgn(|s_{ij}| - \Hat{\beta})\\
=&\; \sgn(s_{ij} + \Hat{\beta} ) \;  \sgn(s_{ij} - \Hat{\beta} )\\
=& \; \sgn(\Delta y_{ij} +\Hat{\beta} \Delta x_{ij}) \;  \sgn(\Delta y_{ij} - \Hat{\beta} \Delta x_{ij})\\
=& \;\sgn(x_i'' - x_j'') \; \sgn(y_i'' - y_j'')
\end{align*}
where $\bx'' = \by + \Hat{\beta} \bx$ and $\by'' = \by - \Hat{\beta} \bx$.
This allows us to compute the influence scores for all observations in the data in $\mathcal{O}(n\log(n))$ time, as stated in the Proposition \ref{prop:4} below. 

\begin{proposition}\label{prop:4}
Given a sample $(\bx,\by) = (x_1,y_1),\ldots,(x_n,y_n)$ of $n$ bivariate data points and a value $m \in \mathbb{R}$. Let  $\bx'' = \by + m \bx$, $\by'' = \by - m \bx$ and $\tau_{ij} = \sgn(x_i'' - x_j'')\sgn(y_i'' - y_j'')$. 
Then the values $\tau_i$, $i \in \{1\ldots n\}$ can together be computed in  $\mathcal{O}(n\log(n))$ time.
\end{proposition}

This speed of calculation is important, as it guarantees that we can use the proposed influence scores for large data sets. We can now use these influence scores to visually identify influential points. This can be done by plotting for example $\tau_i$ versus the indices in the data set or by sub-groups. 
Note that, thanks to the robustness of the ePBr estimator, $\left|\tau_i\right| \leq \frac{n-1}{n(n-1)/2} = \frac{2}{n}$. Therefore, we plot the values $\tilde{\tau}_i \coloneqq n\tau_i/2$, so that they always live on the interval $[-1,1]$.\\

\begin{figure}[ht]
    \centering
    \includegraphics[width = 0.49\linewidth]{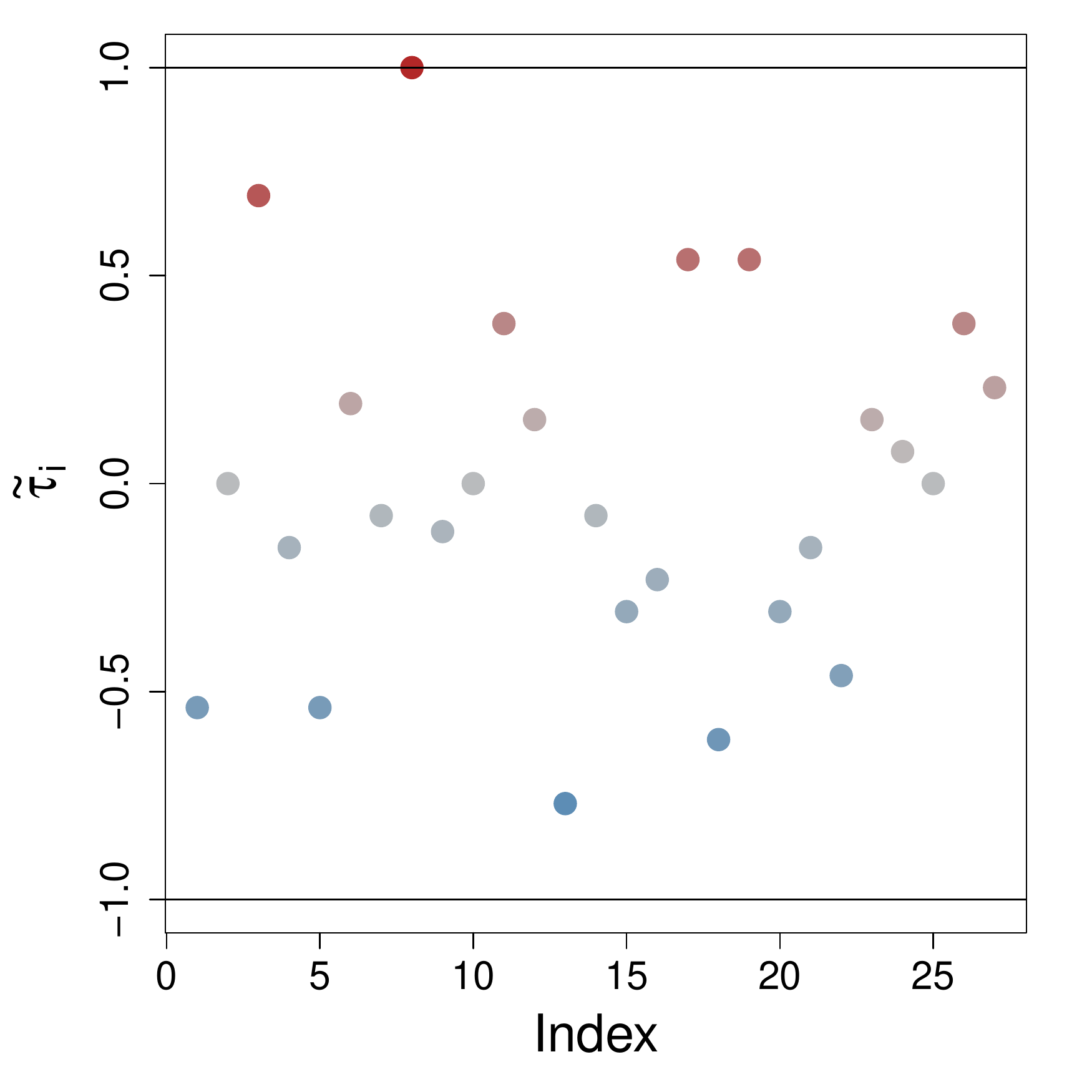}
    \includegraphics[width = 0.49\linewidth]{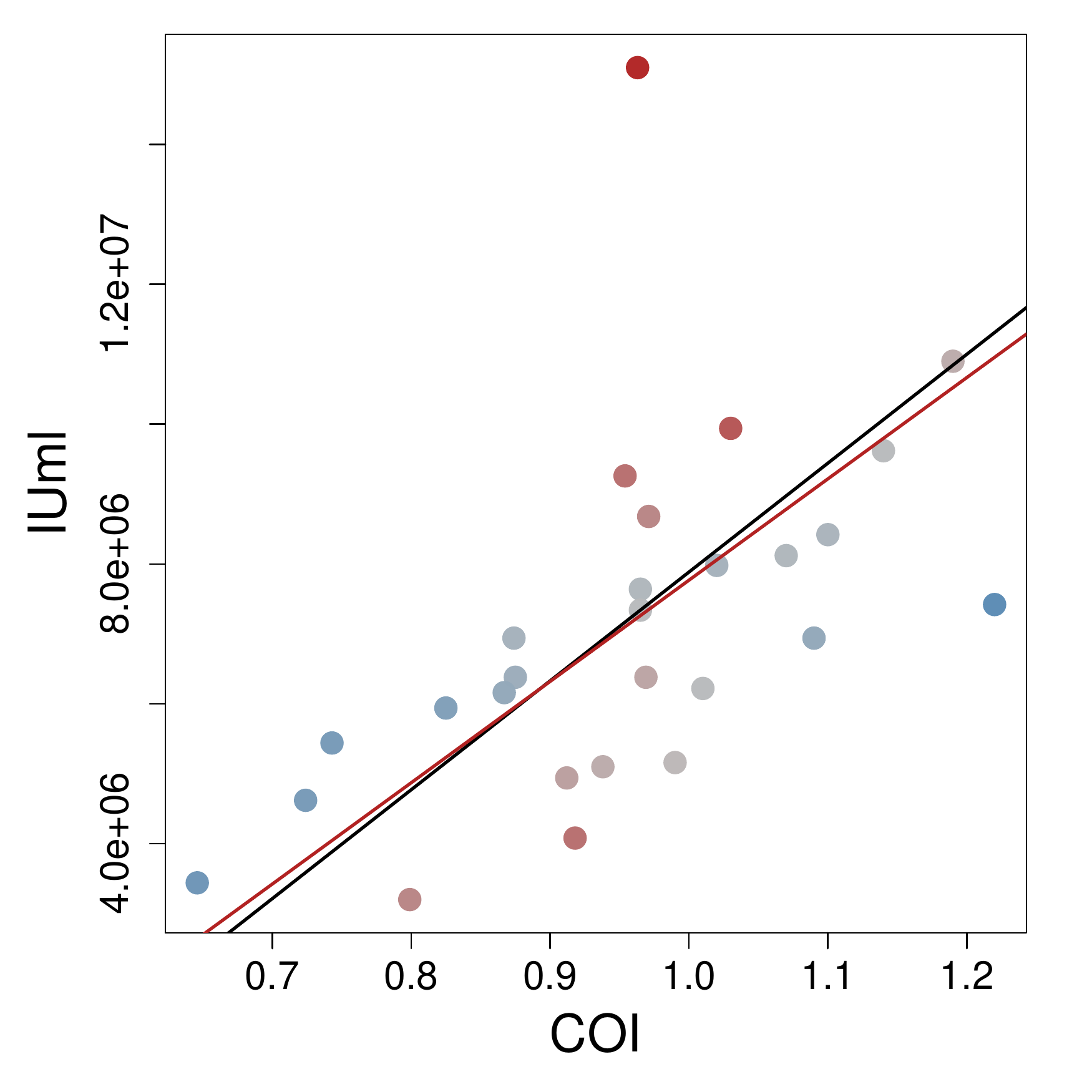}
    \captionsetup{width=\linewidth}
    \caption{A diagnostic plot of the influence score vs. the index (left panel) and a scatter plot of the data (right panel). The right panel also contains the estimated ePBr line in black, as well as the same line when estimated without the highest influence point in red.}
    \label{fig:diagnostic_small}
\end{figure}

As a small example we consider a data set containing 27 observations from a calibration of two clinical chemistry assays, a PCR assay (concentration [IU/ml]) and an immunoassay (cutoff index, COI).
Figure \ref{fig:diagnostic_small} shows the the $\tilde{\tau_i}$ plotted against their index in the left panel. One point (index number 8) immediately stands out as having a high influence score. The right panel of the same figure shows a scatter plot of the data with the points colored according to their value of $\tilde{\tau_i}$. The scatter plot also shows the estimated ePBr line in black, and the same line when estimated without the highest influence point in red. We can clearly see that the single high-influence point attracts the estimated slope, confirming that this observation should be inspected in more detail.

\subsection{Inference}
\label{sec:inference}
In clinical studies, the use of ePBr is most commonly accompanied by inference on the estimated intercept, slope and the bias at medical decision point (see \cite{dufey2020derivation}). This is done either through the use of the theoretical asymptotic distributions of the quantities at interest, or through bootstrapping. We focus on inference on the estimated slope parameter, but the ideas related to bootstrap-based inference can be used for the analysis of the intercept and bias at medical decision point as well.\par

The asymptotic distribution of the estimated slope parameter can be derived using the connection between ePBr and Kendall's rank correlation coefficient (or Kendall's $\tau$) \cite{kendall1938new}. It can be shown that a $1-\alpha$ confidence interval for the slope is given by
\begin{equation}\label{eq:ci_analytical}
    [\tau^{-1}(-z_{1-\alpha/2} \sigma_{\tau}), \tau^{-1}(z_{1-\alpha/2} \sigma_{\tau})]
\end{equation}
where $\tau(m)$ denotes Kendall's correlation between the vectors $\bx'' = \by + m \bx$ and $\by'' = \by - m \bx$ and $\sigma_{\tau}^2$ is the variance of this correlation estimator.\par
The original confidence interval given by \cite{passing1983new} uses $ \sigma_{\tau}^2 = \frac{2(2n+5)}{9n(n-1)}$ as an expression for the variance of Kendall's tau. The conditions for this expression to hold are rather restrictive, which is why a more general expression given by 

\[
    \hat{\sigma}_{\tau}^2 = \frac{n(n-1)\sum_i{\left({\tau_{i}}\right)^2} - 2}{(n-2)(n-3)}
\]
is to be preferred (\cite{daniels1947significance}). As hinted towards in \cite{dufey2020derivation}, this estimate of the variance can be computed in $\mathcal{O}(n\log(n))$ time which follows from the proposition \ref{prop:4}.

As an alternative to working with the asymptotic variance, one can resort to bootstrap-based inference. This is where our fast algorithm really shines, as the repeated estimation of ePBr for multiple bootstrap samples quickly becomes computationally cumbersome. There are multiple ways of obtaining confidence intervals based on bootstrapping. For the sake of simplicity, we will only consider quantile-based confidence interval given by \begin{equation}\label{eq:ci_bootstrap}
[\Hat{Q}_{\alpha/2}(\Hat{\beta}_b), \Hat{Q}_{1-\alpha/2}(\Hat{\beta}_b)]
\end{equation} 
where $\Hat{Q}_{\alpha/2}(\Hat{\beta}_b)$ denotes the $\alpha/2$ empirical quantile of the bootstrapped slopes. For an overview of alternative bootstrapping techniques, we refer to \cite{efron1992bootstrap,efron1994introduction}, and \cite{newson2006confidence}.
Section \ref{sec:examples} contains illustrations of these approaches to inference and highlights where the fast algorithm unlocks new possibilities.



\section{Simulation study}\label{sec:simulation}
In this section we briefly illustrate the computation time of our algorithm and compare it with a naive implementation. In order to do so, we generate $n$ bivariate samples from the model $Y = X + \varepsilon$ where $X \sim \mathcal{N}(0,1)$ and $\varepsilon \sim \mathcal{N}(0, 0.1^2)$. We first compare the proposed algorithm with the naive approach of computing all pairwise slopes and taking the median. This comparison is done on samples of size $n = 10^j$ where $j = 1,\ldots,4$, and mean computation times over 100 replications are shown in Table \ref{tab:comptimes_smalln}. The table illustrates that for small sample sizes, the computation times of the two approaches don't differ too much. For $n = 10^3$, however, there is already a relative difference of order $10^2$ between the two methods. For $n = 10^4$ this gap widens further to a relative difference of order $10^3$.  

\begin{table}[ht]
\centering
\begin{tabular}{llll}
  \hline
n & ePB [$\mu$s] & naive [$\mu$s] \\ 
  \hline
10 & 28 & 803  \\ 
  100 & 210 & 1950 \\ 
  1000 & 3$\times10^3$ & 250$\times10^3$ \\ 
  10000 & 34$\times10^3$ & 34.593 $\times10^6$ \\ 
   \hline
\end{tabular}
 \captionsetup{width=0.8\textwidth}
\caption{Comparison of the mean computation times of the proposed implementation versus a naive implementation of the PB regression estimator. The sample sizes $n$ are equal to $10$, $10^2$, $10^3$ and $10^4$.} 
\label{tab:comptimes_smalln}
\end{table}

\begin{figure}[!h]
    \centering
    \includegraphics[width = 0.7\linewidth]{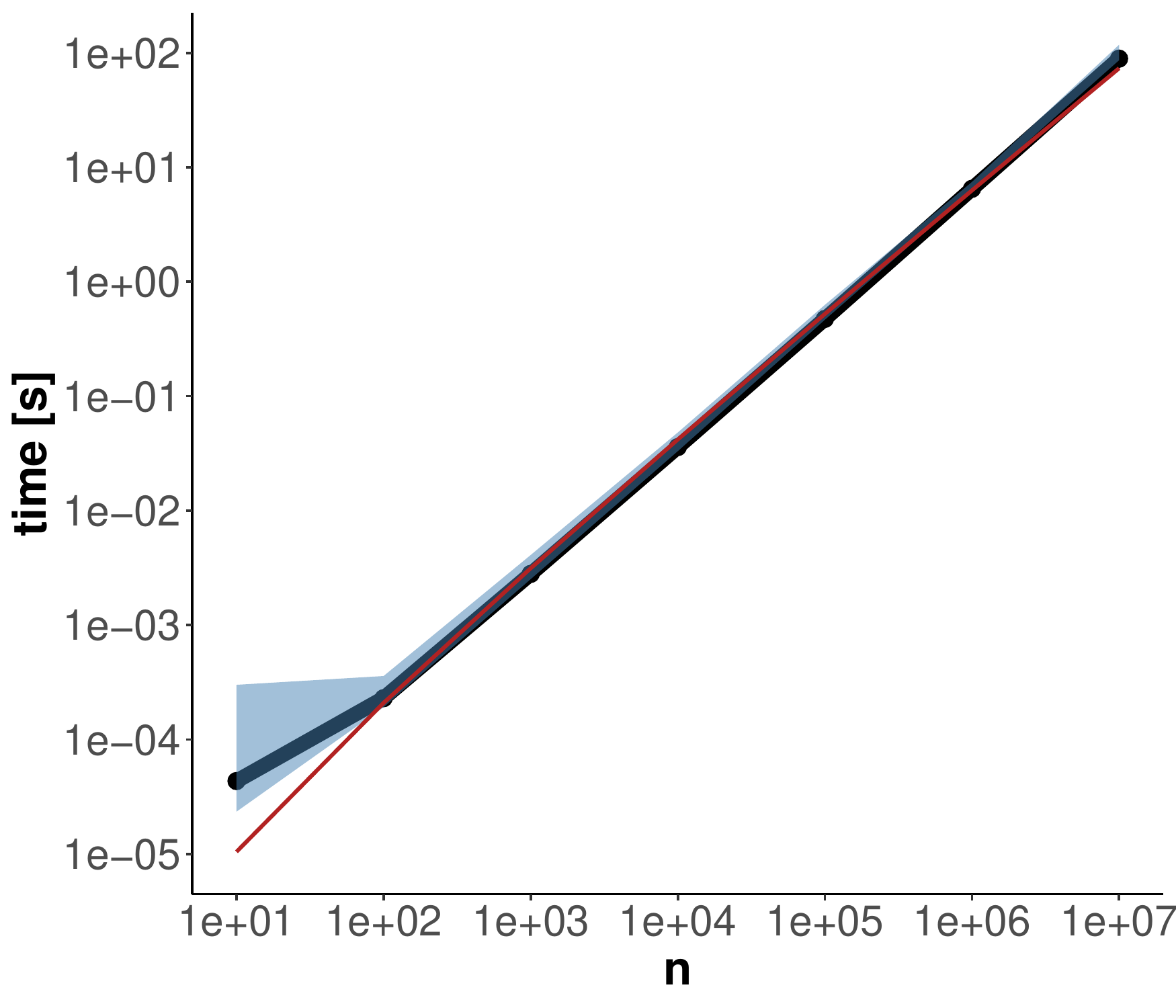}
  \captionsetup{width=0.7\linewidth}
    \caption{Average computation time in seconds for the fast equivariant Passing--Bablok algorithm for 100 replicates. The blue shades indicate the minimum and maximum time, whereas the red line is an estimate of the $\sim n \log(n)$ computational cost.}
    \label{fig:comptimes_largen}
\end{figure}

A naive implementation is not practical for $n > 10^4$. Therefore, we continue the simulation study for $n = 10^5, 10^6$, and $10^7$ on the proposed fast estimator only. The result of this is shown in Figure \ref{fig:comptimes_largen}, where the blue shade indicates the minimum and maximum computation times and the black line indicates the mean computation time. It is clear that the relative variance in computation time is larger for very small values of $n$, but this is mainly due to the computational overhead and it disappears quickly for increasing sample size. The additional red line shows the function $f(n) = c\,n \log(n)$ where $c = 4.5\times10^{-7}$ s, indicating that the computation time does indeed grow at the expected $\mathcal{O}(n\log(n))$ rate.

\section{Real data example}\label{sec:examples}
We now analyze a data set containing the result of a comparison of 55808 patient samples with two clinical tests performed in four different laboratories. This data set is large enough such that the naive computation of the ePB estimate, let alone bootstrap based inference, is infeasible on most computer architectures. The proposed algorithm however, deals with this data without any issues.\\

\begin{figure}[ht]
    \centering
    \includegraphics[width = 0.49\linewidth]{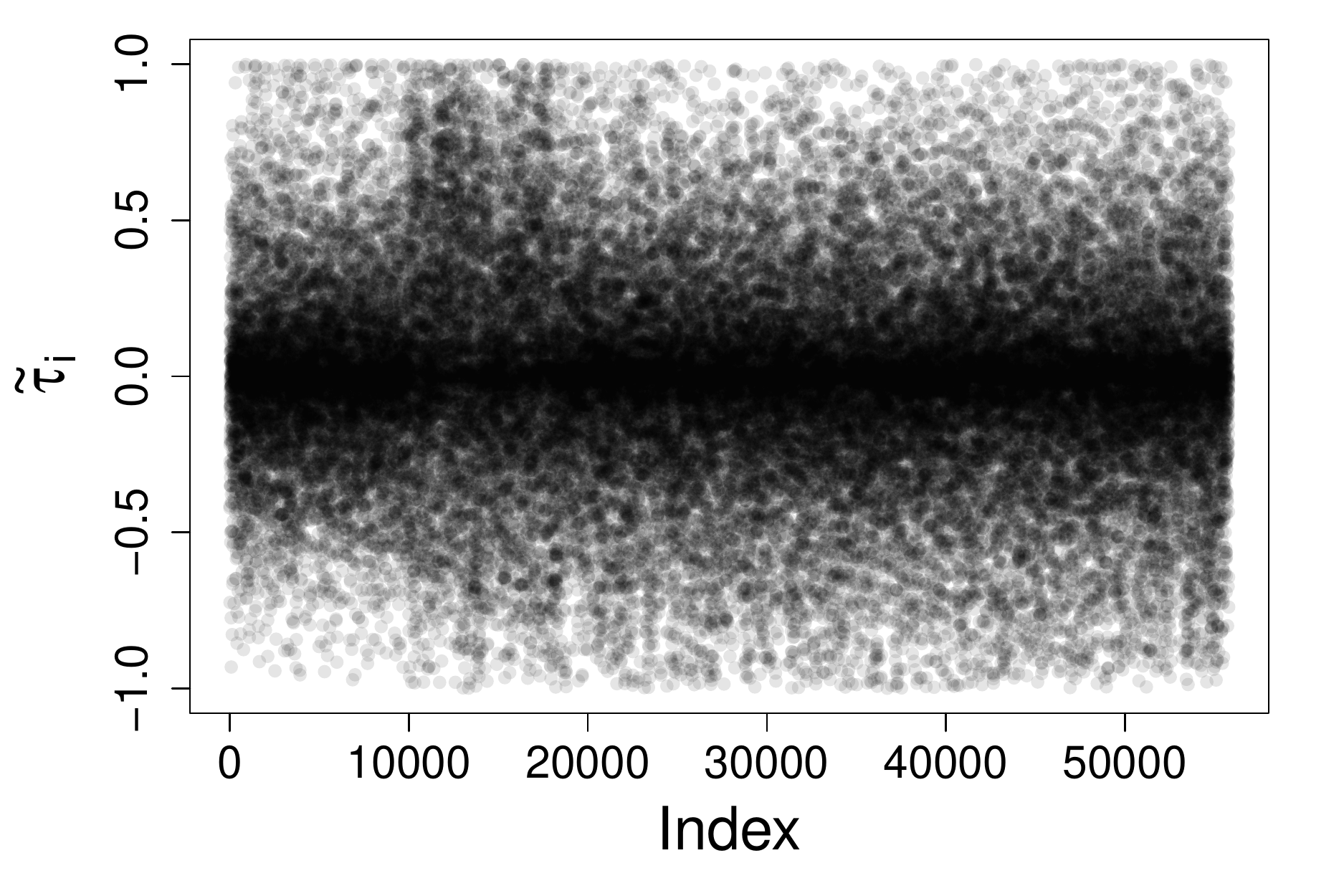}
    \includegraphics[width = 0.49\linewidth]{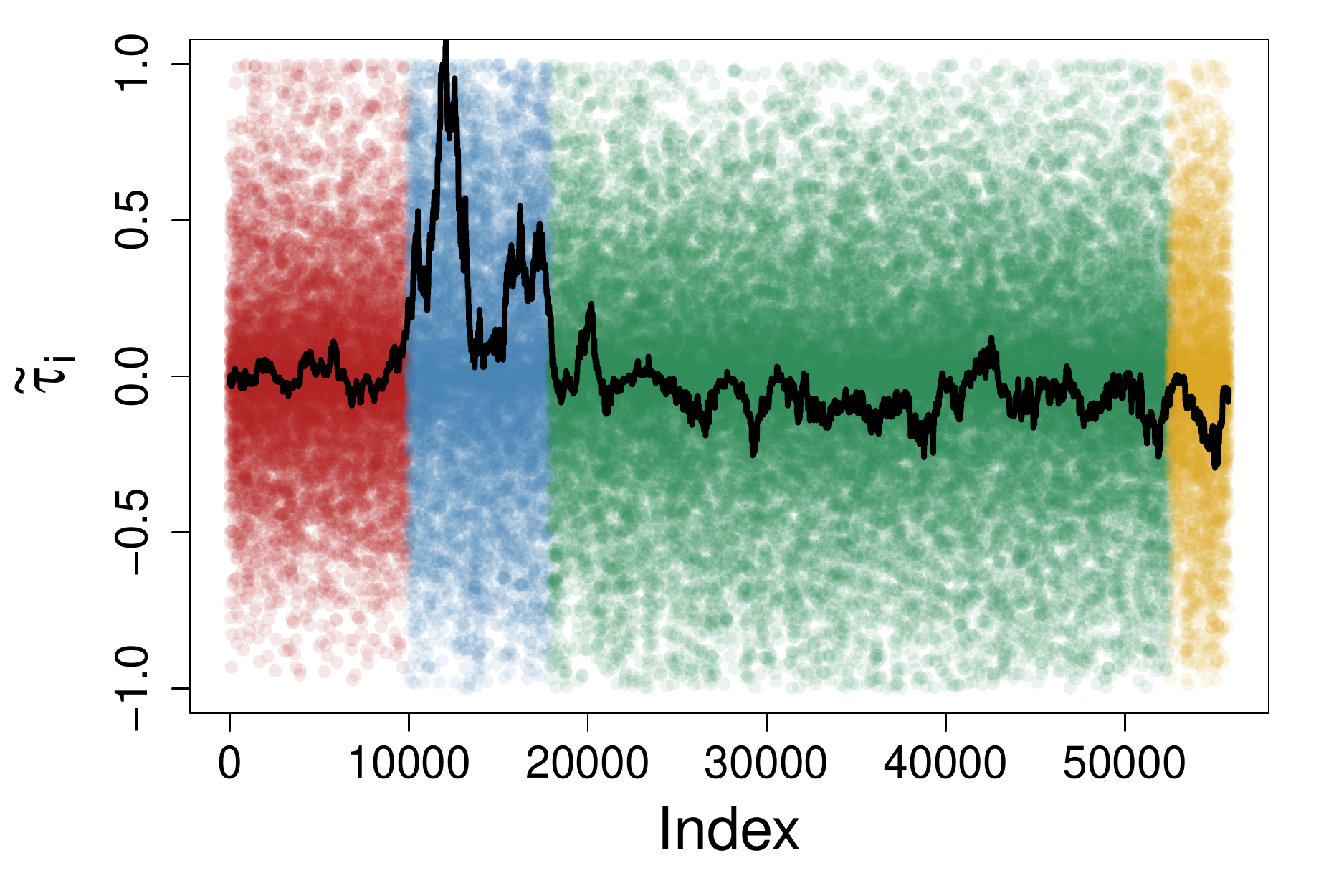}
    \captionsetup{width=\linewidth}
    \caption{The scaled influence scores plotted against the index (left), and the same plot with an estimation of the trend in black (right). The different colors in the right panel pertain to different testing centers.}
    \label{fig:diagnostic}
\end{figure}

We first use the proposed diagnostic tool to identify influential points. Computing the influence scores for all observations in the data takes less than one second. Figure \ref{fig:diagnostic} shows a plot of the (scaled) influence scores against the indices in the data. The left panel shows the plot without any additional markings. This plots suggests that there is a concentration of influential points between index 10 000 and 20 000 in the data. It turns out that this patch of tests corresponds to tests taken in a given testing center. This is illustrated in the right panel of Figure \ref{fig:diagnostic}, where the points are shown in the color of their testing center. The black line in this figure presents a trend based on a moving average, again confirming that there is a clear concentration of influential points in the blue testing center. The influence scores thus identify a potential irregularity with comparison tests coming from this testing center, as the observations from this center strongly pulled the estimated slope upwards. 


Having identified that the 4 different test centers should probably be treated separately, we proceed with the inferential analysis. As described in section \ref{sec:methodology}, we can construct confidence intervals based on the asymptotic distribution as in Equation \ref{eq:ci_analytical} or based on the bootstrap as in Equation \ref{eq:ci_bootstrap}. Table \ref{tab:inference} shows the results of this analysis. 
While the calculation of point estimates was up to now practically prohibitive for data sets of this size,with the present algorithm, even bootstrap confidence intervals based on 1000 bootstrap samples can be calculated in less than half an hour. 
The analytic and bootstrap confidence intervals coincide very well, which supports the validity of the confidence intervals developed in sec.\ \ref{sec:inference} and in \citet{dufey2020derivation}. 
Hence, for even larger data sets where bootstrapping is still demanding, one can rely on analytical asymptotic confidence intervals.

\begin{table}[ht]
\centering
\begin{tabular}{rrrrrrrr}
  \hline
 & $n$& $\hat{\beta}$ &$\mathrm{LCI}_\mathrm{b}$ & $\mathrm{UCI}_\mathrm{b}$ & $\mathrm{LCI}_\mathrm{a}$ & $\mathrm{UCI}_\mathrm{a}$ & time (s) \\ 
  \hline
ALL & 55808 & 0.9939 & 0.9920 & 0.9959 & 0.9917 & 0.9960 & 1462 \\ 
  A & 9996  & 1.0038 & 1.0000 & 1.0084 & 0.9996 & 1.0080 & 209 \\ 
  B & 7929  & 1.0291 & 1.0230 & 1.0352 & 1.0230 & 1.0352 & 160 \\ 
  C & 34600 & 0.9833 & 0.9808 & 0.9854 & 0.9809 & 0.9856 & 935 \\ 
  D & 3285  & 0.9734 & 0.9640 & 0.9822 & 0.9641 & 0.9827 & 65 \\ 
   \hline
\end{tabular}
\caption{Sample sizes, point estimates, bootstrap confidence intervals (denoted with subscript b), analytical confidence intervals (denoted with subscript a) and computation time (for the bootstrap CIs) for the slope parameter. The values are given for the full dataset, as well as for each individual test center A to D.} 
\label{tab:inference}
\end{table}

\section{Conclusion}\label{sec:conclusion}
We have proposed a computationally efficient algorithm for the equivariant Passing--Bablok regression estimator. It allows for the exact computation of the estimator on much larger data sets than previously possible. An additional benefit of the fast algorithm is the applicability to bootstrap based inference. Finally, we have proposed a diagnostic tool for identifying possible outlying  points in ePB regression, which can also be calculated quickly. We have illustrated the proposed tools on a real data set containing 55808 comparison tests. Thanks to the achieved computational speedups, it is now very easy to analyze these data and the proposed influence scores prove very helpful to identify irregularities e.g.\ with respect to the testing centers.



\clearpage
\bibliographystyle{chicago}
\bibliography{ePBR_bib}

\clearpage

\appendix
\section*{Supplementary material}
Supplementary material with proofs is available upon request.

\end{document}